# Photon-statistics force in ultrafast electron dynamics

**Matan Even Tzur[†,1], Michael Birk[†,1,2], Alexey Gorlach[2], Michael Krüger[1], Ido Kaminer[2] and Oren Cohen[1]**

[1] Solid State Institute and Physics Department, Technion – Israel Institute of Technology, Haifa 3200003, Israel

[2] Solid State Institute and Department of Electrical and Computer Engineering, Technion – Israel Institute of Technology, Haifa 3200003, Israel.

In strong-field physics and attosecond science, intense light induces ultrafast electron dynamics. Such ultrafast dynamics of electrons in matter is at the core of phenomena such as high harmonic generation (HHG), where these dynamics lead to emission of extreme UV bursts with attosecond duration. So far, all ultrafast dynamics of matter were understood to originate purely from the classical vector potential of the driving light, disregarding the influence of the quantum nature of light. Here we show that dynamics of matter driven by bright (intense) light significantly depend on the quantum state of the driving light, which induces an effective photon-statistics force. To provide a unified framework for the analysis & control over such a force, we extend the strong-field approximation (SFA) theory to account for non-classical driving light. Our quantum SFA (qSFA) theory shows that in HHG, experimentally feasible squeezing of the driving light can shift & shape electronic trajectories and attosecond pulses at the scale of hundreds of attoseconds. Our work presents a new degree-of-freedom for attosecond spectroscopy, by relying on nonclassical electromagnetic fields, and more generally, introduces a direct connection between attosecond science and quantum optics.

# INTRODUCTION

Ultrafast electron dynamics driven by intense light is central to all phenomena in strong-field physics and attosecond science[1]. Examples include attosecond interferometry[2–4], attosecond band gap dynamics[5], ultrafast chiral detection[6–8], and more [9,10,11]; the prime example being high harmonic generation (HHG)[12–16]. HHG occurs when an intense light field drives matter (gases[12–14], liquids[17], solids[18], or plasma[19]) to emit a train of attosecond pulses[20], which spectrally consists of high-order harmonics of the driving field. HHG in gas phase is intuitively understood in terms of the three-step model[14–16]: an initially bound electron undergoes laser-induced tunnel ionization, then accelerates in the continuum by the oscillating laser field, and finally recombines with its parent ion, releasing its kinetic and potential energy as a high-energy photon.

Soon after its inception, the classical three-step model[15] was generalized by *Lewenstein et al.* to a quantum mechanical setting[16], where matter is quantum and light is classical. Lewenstein's model, also known as the strong-field approximation (SFA) theory, allows for both quantitative analysis and intuitive understanding of the attosecond pulse generation of HHG in terms of ultrafast electron dynamics. In particular, the duration and chirp of the attosecond pulses are directly correlated with electronic recombination times, and indirectly correlated with ionization times and canonical momenta of the electronic trajectories. Indeed, the SFA theory plays a critical role in the development of attosecond science[21,22,23].

Nevertheless, so far, the SFA theory remained semi-classical – it did not account for the quantum nature of the driving light. The reason for this was twofold. First, since the driving pulse contains a macroscopic number of photons, its quantum nature was expected to be insignificant, i.e., only the classical vector potential of the light field played any role in the induced dynamics. Second, all experiments in strong-field physics and attosecond science used classical driving fields (i.e., Glauber coherent-states[24]), as it was thought to be the only state of light accessible with high intensities and ultrashort pulse durations. Importantly, this situation is now rapidly changing, because ultrashort pulses of intense non-classical light are becoming well-established drivers of nonlinear optics[25,26], starting to approach the regime of strong-field physics[25,27–31]. This rapid advance motivates us to revisit the established SFA theory and consider the effect of the quantum photon statistics of the driving light on the ultrafast dynamics of the electrons. Insight into the dynamics of electrons in bright quantum fields is especially timely considering the recent developments in quantum optical strong-field physics. Many years after the early quantum-electrodynamical investigations of HHG[32,33], this field is reignited by recent photon-counting experiments[35–37], and it continues to develop as a platform for quantum information processing, e.g., through generation of photonic cat-states[34], and more[35–39].

Here we present the framework of quantum-SFA, extending Lewenstein's SFA theory to account for non-classical states of intense light driving atoms or other emitting systems[40]. The framework of qSFA reveals a new effect: ultrafast dynamics of strongly light-driven matter significantly depends on the quantum state of the driving light, and specifically on its photon statistics. We study this effect in detail for the example of squeezed light and show that experimentally feasible squeezing of the driver's photonic state modifies HHG ionization and recombination times by hundreds of attoseconds. Moreover, this modulation of electronic trajectories by squeezing leaves a pronounced mark on the temporal profile of the emitted attosecond pulses and on the HHG spectrum. We further show how to explain these phenomena in terms of the short and long trajectories. Finally, we interpret our results in terms of an effective photon-statistics force, which is created by the squeezed pump and applies on the electronic system. In general, the qSFA framework opens the way for the fusion of quantum optics with attosecond science, and for time-resolved insight about the role of the quantum nature of light in the interplay between three of the most fundamental processes in physics – ionization, recombination, and acceleration of charged particles by photons.

**RESULTS**

**The quantum-optical SFA**

A main result of the semi-classical SFA theory is an analytical formula for the dipole moment expectation value of an atom in bright & classical laser light. This result played a central role in the development of attosecond science over the years, as it connects between the HHG emission and the underlying electronic dynamics. In this section, we generalize this result and derive an analytical formula for the dipole moment expectation value for an electron driven by an arbitrary quantum state of light. To do so, we decompose the quantum state of the driving light to Glauber (semi-classical) coherent states, where each coherent state is denoted by $|\alpha\rangle$. We find that within the SFA, the dipole moment expectation value $z(t)$ of an electron driven by such a superposition state is given by $z(t) = \int d^2\alpha \, P(\alpha) z_\alpha(t)$ where $P(\alpha)$ is a distribution of coherent states specifying the quantum state of the driver, and $z_\alpha(t)$ are semi-classical dipole moments, corresponding to an electron driven by a coherent state $|\alpha\rangle$.

We begin by considering an atomic system that interacts with light of an arbitrary quantum state that can be approximated as a single mode during the interaction process. The ionization potential of this atomic system is denoted by $I_p$. The density matrix of the driving light is specified by the *generalized P representation*[41] $P(\alpha, \beta)$:

$$\hat{\rho}_{drive} = \int d^2\alpha \, d^2\beta \, P(\alpha, \beta) \frac{|\alpha\rangle\langle\beta^*|}{\langle\beta^*|\alpha\rangle}. \quad (1)$$

$|\alpha\rangle = |\alpha_x + i\alpha_y\rangle$ is a coherent state of light, where $\alpha_x$ and $\alpha_y$ are real-valued parameters that correspond to a classical electromagnetic wave whose vector potential is given by[42]

$$\mathbf{A}_\alpha(t) = \frac{\epsilon^{(1)}}{\omega}\left(\alpha e^{-i\omega t} + \alpha^* e^{i\omega t}\right)\hat{z} = \frac{2\epsilon^{(1)}}{\omega}\left[\alpha_x \cos(\omega t) + \alpha_y \sin(\omega t)\right]\hat{z}. \quad (2)$$

Here, $\omega$ is the angular frequency of the field, $\epsilon^{(1)} = \sqrt{\hbar\omega/2\epsilon_0 V}$ is the so-called *single-photon-amplitude*[42], and $V$ is the volume of quantization, which will be eventually eliminated through the limit $V \to \infty$. The polarization is linear along $\hat{z}$. The corresponding electric field is given by $\mathbf{E}_\alpha(t) = -(\partial \mathbf{A}_\alpha/\partial t) \cdot \hat{z} = 2\epsilon^{(1)}\left[-\alpha_x \sin(\omega t) + \alpha_y \cos(\omega t)\right]\hat{z}$.

The density matrix of an electron driven by $\hat{\rho}_{drive}$ is (SI section I):

$$\hat{\rho}_e = \int d^2\alpha \, d^2\beta \, P(\alpha, \beta^*) |\phi_\alpha(t)\rangle\langle\phi_\beta(t)|, \tag{3}$$

where $|\phi_{\alpha,\beta}\rangle$ are solutions to the semi-classical time-dependent Schrodinger equation (TDSE)

$$i\hbar \frac{\partial |\phi_{\alpha,\beta}(t)\rangle}{\partial t} = \left[-\frac{1}{2m}\nabla^2 + U(z) - z \cdot E_{\alpha,\beta}(t)\right] |\phi_\alpha(t)\rangle. \tag{4}$$

$U(z)$ is the atomic potential. Most generally, the dipole moment expectation value is given by

$$z(t) = \text{Tr}[\rho_e(t)x] = \int d^2\alpha \, d^2\beta \, P(\alpha, \beta^*) \langle\phi_\beta(t)|z|\phi_\alpha(t)\rangle \tag{5}$$

Neglecting depletion of the ground state, keeping only bound-continuum transitions[16] and utilizing the property $P(\alpha, \beta) = P(\beta, \alpha^*)$ of the generalized P representation, we may write the dipole moment expectation value as (SI section II)

$$z(t) = \int d^2\alpha \, P(\alpha) z_\alpha(t), \tag{6}$$

in which $z_\alpha(t)$ is the dipole moment expectation value for a coherent-state drive $|\alpha\rangle$, and $P(\alpha) = \int d^2\beta \, P(\alpha, \beta^*)$. This $P(\alpha)$ is not to be confused with the Glauber-Sudarshan $P$ representation (SI section II). Eq. (6) quantifies the dipole for a driving field with arbitrary photon statistics. This dipole is later used to derive the HHG spectrum and pulse shape and is therefore a central result of this work. Finally, we take the limit where the single-photon amplitude $\epsilon^{(1)}$ approaches zero (i.e., infinite quantization volume $V$), while keeping the intensity (photon density) of the driving photonic state constant. In this limit, $P(\alpha)$ of a coherent state reduces to a Dirac delta function (SI section III).

**High harmonic generation driven by squeezed coherent light**

The previous section formulates how the quantum nature of light alters dynamics in light-matter interactions. Notably, under the strong-field approximation, the main effect on the dynamics is due to photon statistics. Qualitatively, this means that one could have two driving fields carrying the same classical vector potential but differing in photon statistics resulting in different material dynamics. In this section, we further investigate this effect, focusing on squeezed coherent (SC) driving light fields. An SC field is denoted by $|\gamma = \gamma_x + i\gamma_y, r\rangle$, with $r$ denoting the degree of squeezing and $\gamma$ representing the coherent-state part that has a vector potential $A_\gamma(t)$ (Eq. (2)). The photon number of the SC field is $N_{|\gamma,r\rangle} = |\gamma|^2 + \sinh^2(r)$. Such a field can be generated by combining in a beam splitter a strong coherent-state beam with a comparatively weak (~1%) squeezed-vacuum beam[43]. The squeezing phase may be tuned by adjusting the phases between the coherent and squeezed-vacuum arms (Figure 1). The squeezing parameter $r$ is set (without loss of generality) to be real and positive henceforth, i.e., the $x$ quadrature is squeezed.

Inserting the $P(\alpha)$ distribution associated with $|\gamma, r\rangle$ into Eq. (6) and changing the integration variables to electric field quadratures $E_{\alpha_x} = 2\epsilon^{(1)}\alpha_x$, $E_{\alpha_y} = 2\epsilon^{(1)}\alpha_y$, leads to an expression for the dipole moment induced by the interaction of an atom with such a SC state (SI section III):

$$z_{\text{SC}}(t) = \frac{1}{\sqrt{2\pi}|E_{\text{vac}}|} \int dE_{\alpha_y} e^{-\frac{(E_{\alpha_y} - E_{\gamma_y})^2}{2|E_{\text{vac}}|^2}} z_{E_{\alpha_y}, E_{\gamma_x}}(t), \quad (7.\text{a})$$

$$I_{\text{vac}} \equiv \frac{1}{2}\epsilon_0 c |E_{\text{vac}}|^2 \equiv c\hbar\omega \frac{\sinh^2(r)}{V}, \quad (7.\text{b})$$

where $z_{E_{\gamma_x}, E_{\alpha_y}}(t)$ is the dipole moment expectation value induced by a classical driving field $2\epsilon^{(1)}[-\gamma_x \sin(\omega t) + \alpha_y \cos(\omega t)]$, and we have defined $I_{\text{vac}}$ and $E_{\text{vac}}$, the intensity and amplitude of the squeezed vacuum part of the SC state. The total intensity of the SC beam is given by $I_{\text{tot}} = I_{\text{coh}} + I_{\text{vac}}$, where $I_{\text{coh}} = c\hbar\omega|\gamma|^2/V$ stems from the classical electromagnetic field of the beam and $I_{\text{vac}}$ from squeezing.

Eq. (7.a) allows to calculate the HHG spectrum and the underlying attosecond pulses driven by squeezed state of light. We evaluate the dipoles $z_{E_{\alpha_y}, E_{\gamma_x}}(t)$ by solving Eq.(4) numerically for a model Ne atom[44]. The resulting attosecond pulses and HHG spectra are depicted in Figure 1 (all cases presented exhibit $I_{\text{coh}} = 2 \times 10^{14}\,\text{W/cm}^2$ and $I_{\text{vac}} = 2 \times 10^{12}\,\text{W/cm}^2$). We find that a phase-squeezed driver creates horn-shaped attosecond pulses, and smeared HHG spectra that exhibit a less pronounced exponential cutoff due to increased amplitude uncertainty. In contrast, an amplitude-squeezed driver creates pulse-to-pulse envelope fluctuations in the attosecond pulses, and non-integer spectral peaks in the HHG spectra due to increased phase uncertainty.

(a) Schematic illustration of high harmonic generation (HHG) with squeezed light

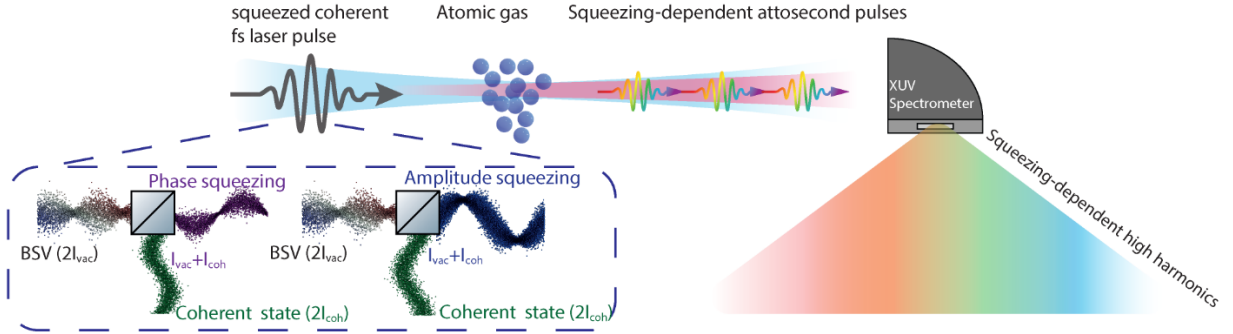

(b) Attosecond pulses & HHG driven by coherent, phase-squeezed, and amplitude-squeezed light

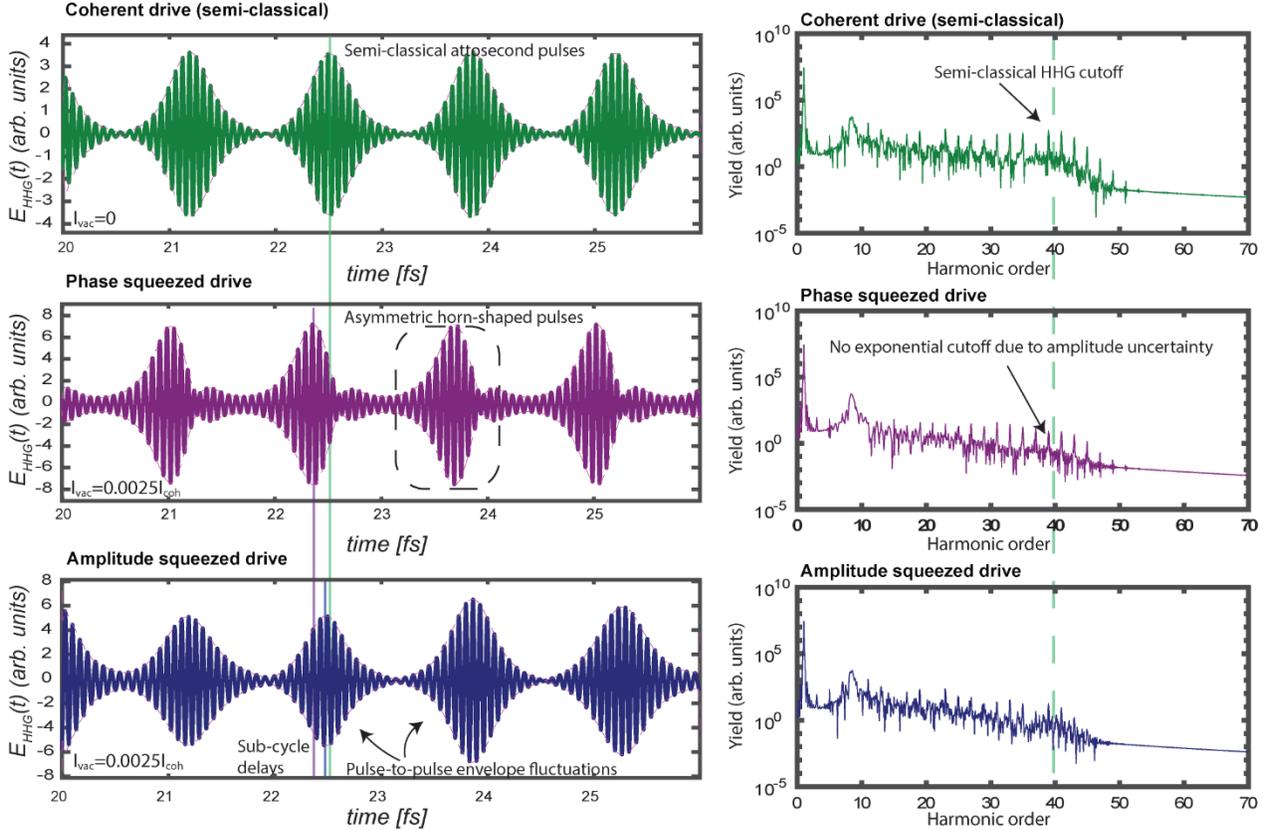

**Figure 1: Squeezing-dependent generation of attosecond pulses and high-order harmonics. (a)** Schematic illustration of high harmonic generation (HHG) with a squeezed light state. **(b)** Exemplary attosecond pulses and HHG spectra driven by (top) green: classical light, i.e., coherent state $|\alpha, 0\rangle$, (center) purple: phase-squeezed coherent state $|\alpha, r\rangle$, and (bottom) blue: amplitude-squeezed coherent state $|\alpha, -r\rangle$. The plots in all panels of both figures are based on Eq. (7.a), choosing a squeezing parameter $r$ that satisfies $sinh^2(r) = 0.0025|\alpha|^2$, where $\alpha$ is normalized to correspond to an average intensity of $2 \times 10^{14}\ W/cm^2$. Explicitly, The parameters we choose for the phase-squeezed driving light are $E_{\gamma_y} = 0.0756$ a.u., $E_{\gamma_x} = 0$, $E_{vac} = \sqrt{0.0025}E_{\gamma_y}$, and for the amplitude-squeezed driving light are $E_{\gamma_y} = 0$, $E_{\gamma_x} = 0.0756$ a.u., $E_{vac} = \sqrt{0.0025}E_{\gamma_x}$.

**Electron trajectories in quantized fields**

The previous section shows that indeed, in HHG, the emitted attosecond pulses and spectrum are sensitive to the photon statistics of the driving field, and this is reflected in the sensitivity of the underlying material dynamics through the dipole moment expectation value. Therefore, one also expects the underlying electronic trajectories to be sensitive to photon statistics, even without changing the mean instantaneous electric field of the driver. To quantify the sensitivity of the electronic trajectories to photon statistics, we now derive coupled equations[45] for short & long electronic trajectories in a squeezed coherent laser field. We solve these equations and show that the electronic trajectories and attosecond pulses are strongly modified by squeezing of the driver's photonic state.

To solve for the electronic trajectories induced by the quantum field, we plug into Eq.(6) the semi-classical SFA dipole moments $z_\alpha(t)$[16]:

$$z_\alpha(t) = \frac{i}{\hbar}\int d^3\boldsymbol{p} \int_0^t dt' E_\alpha(t) d_z(\boldsymbol{p} - \boldsymbol{A}_\alpha(t')) d_z^*(\boldsymbol{p} - \boldsymbol{A}_\alpha(t)) \cdot e^{-\frac{i}{\hbar}S_\alpha(\boldsymbol{p},t',t)} + \text{c. c.}, \quad (8.\text{a})$$

$$S_\alpha(\boldsymbol{p}, t', t) \equiv \int_{t'}^t dt'' \left(\frac{[\boldsymbol{p} - \boldsymbol{A}_\alpha(t'')]^2}{2} + I_p\right). \quad (8.\text{b})$$

Here, $\boldsymbol{p} = \boldsymbol{v} + \boldsymbol{A}_\alpha(t)$ [16], $d_z(\boldsymbol{v}) = \langle\boldsymbol{v}|z|0\rangle$ where $|0\rangle$ is the ground state of the atomic system, $|\boldsymbol{v}\rangle$ is a continuum state of velocity $\boldsymbol{v}$, and the function $S_\alpha(\boldsymbol{p}, t, t')$ is the semi-classical action of an electron driven by a coherent state $|\alpha\rangle$. Substituting Eq. (8.a) into Eq. (6) enables us to extract semi-analytical conclusions about the trajectories. Specifically, we can generalize the Lewenstein approach[16] that relies on the saddle point approximation to extract three physically meaningful (and experimentally observable[46]) stationary parameters, $[\boldsymbol{p}, t' = t_0, t = t_1]$, defining each electronic trajectory. Each frequency component is associated with two trajectories (two sets of stationary parameters), corresponding to the short & long trajectories. Physically, $\boldsymbol{p}$ is inferred as the canonical momentum of a trajectory, $t_0$ is the time of ionization, and $t_1$ is the time of recombination.

Plugging Eq.(8.a) into Eq.(6), we find that in the quantum optical case, the semi-classical action $S_\alpha(\boldsymbol{p}, t', t)$ is replaced with a quantum-optical action $S_q(\boldsymbol{p}, t', t, \alpha)$:

$$\underbrace{S_{\text{q}}(\boldsymbol{p},t',t,\alpha)}_{\text{quantum-optical action}} = \underbrace{S_{\alpha}(\boldsymbol{p},t',t)}_{\text{semi-classical action}} + \underbrace{i\log(P(\alpha))}_{\text{photon statistics}}. \tag{9}$$

Similarly, to the semi-classical case, emission of high-order harmonics at frequencies $\Omega = n\omega$ will mainly originate from the stationary points of $S_{\text{q}}(\boldsymbol{\kappa}_{\text{q}}) - \hbar\Omega t$ with respect to $\boldsymbol{\kappa}_{\text{q}} = [p,t',t,\alpha]$. Notably, in the quantum optical picture, each trajectory is characterized by a fourth stationary parameter, which we infer as the dominant coherent component ($\alpha = \alpha_x + i\alpha_y$) associated with the trajectory (SI, section IV). Figure 2 presents the trajectories for a coherent state ($r = 0$, green), phase squeezed state ($r > 0$, purple), and amplitude squeezed state ($r < 0$, blue). The solutions for the stationary parameters are displayed in Figure 2(c). Despite all these states differing only in quantum statistics and having identical coherent electric fields, their electronic trajectories display significant qualitative & quantitative shifts. To explore the relevance of these stationary solutions to the complete solution of HHG driven by squeezed coherent light, we perform a time-frequency analysis (Gabor transform) over the dipole accelerations obtained by Eq.(6) (Figure 2(b)). Each Gabor transform is overlaid with the real parts of the recombination times obtained from the stationary parameters of our qSFA theory, showing good agreement between the plots.

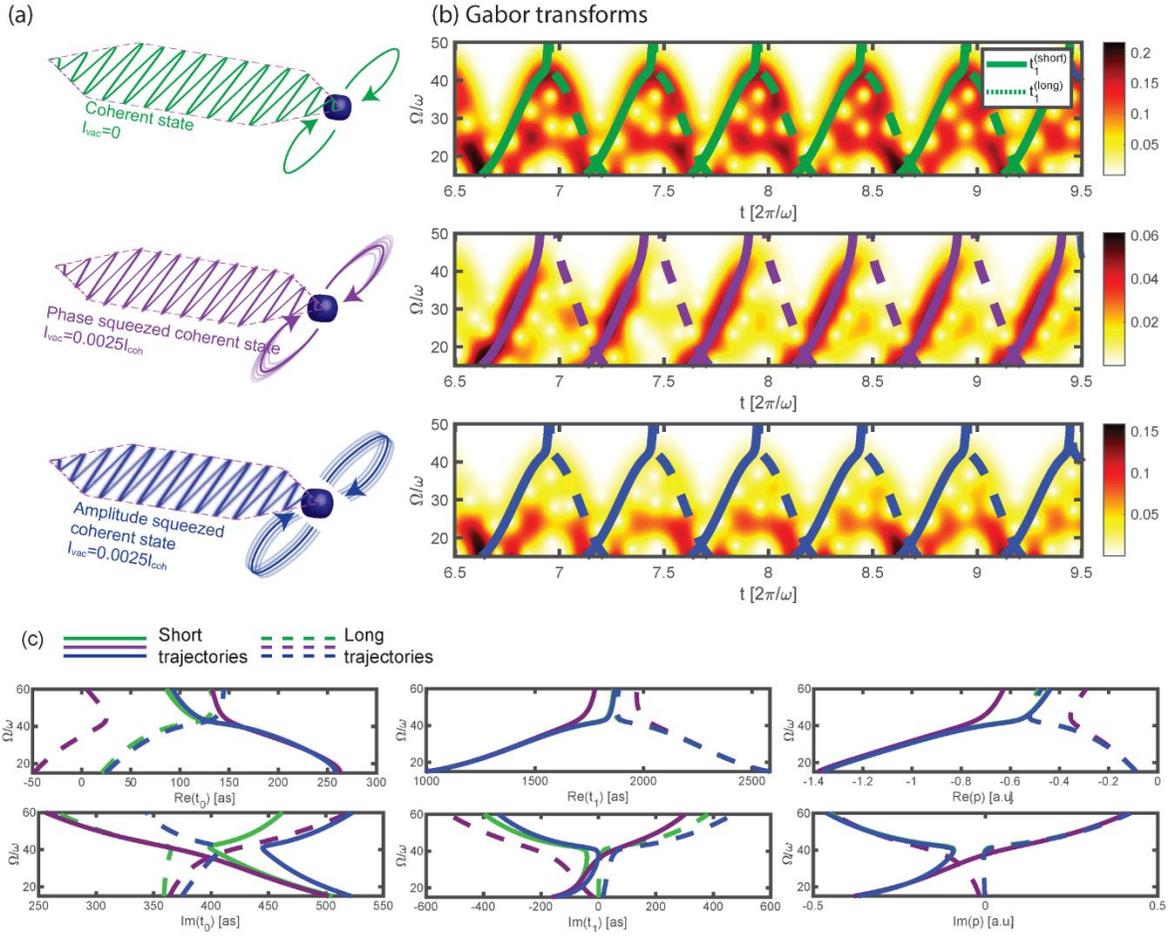

**Figure 2: The three-step model in a quantized field and the resulting electronic trajectories. (a)** Illustration of the trajectories under amplitude and phase uncertainty induced by phase and amplitude squeezing, respectively. **(b)** Time frequency analysis (Gabor transform) of the dipole acceleration of an atom driven by the light states discussed above, overlaid with the real parts of the recombination times of short and long trajectories. **(c)** Solutions of the stationary parameters $[p, t_1, t_0]$ for the cases of coherent state (green), phase squeezing (purple), and amplitude squeezing (blue). The same parameters coherent and squeezing parameters as Figure 1 are used here.

**Time delays of the attosecond pulses and the underlying electronic trajectories**

Having established a connection between photon statistics, attosecond pulses, and electronic trajectories, we now turn to exploring the dependence of the emitted attosecond pulses on the strength of the squeezing. This dependence exhibits a complex structure that we explain in terms of short and long trajectories in the quantized fields, as formulated in the previous section. Using Eq. (7.a), we calculate the attosecond pulses generated by the interaction of a model Ne atom with a squeezed state of light, scanning the range $-0.01 < \text{sign}(r)\, I_{\text{vac}}/I_{\text{coh}} < 0.01$, where in the following, we drop the factor $\text{sign}(r)$ for brevity. Here, $r < 0$ represent amplitude squeezing, $r > 0$ phase squeezing, and $r = 0$ corresponds to a coherent state.

The resulting attosecond pulses are depicted in Figure 3(a). The dashed gray lines mark the position of the peak of the attosecond pulse, $t_{\text{peak}}$, for the case of a coherent pulse. The faded red line follows $t_{\text{peak}}(I_{\text{vac}})$ for the explored range and is shown by itself in Figure 3(b). Notably, $t_{\text{peak}}$ exhibits a strong and complicated dependence on $I_{\text{vac}}$. This dependence can be understood in terms of a competition between recombination times of short and long trajectories, which are displayed in Figures 3(c-d). For $|I_{\text{vac}}|/|I_{\text{coh}}| > 0.5\%$, long trajectories display the most significant shifts in recombination times, hence, they mostly dominate the dependence of $t_{\text{peak}}$ on $I_{\text{vac}}$. For the range $|I_{\text{vac}}|/|I_{\text{coh}}| < 0.1\%$, the short trajectories display the greater shifts, hence, the peak of the attosecond pulse is dominated by their recombination times. Finally, in between these ranges, short and long trajectories display comparable shifts in recombination times, resulting in a complicated curve for $t_{\text{peak}}(I_{\text{peak}})$. Figure 3(e) shows $t_{\text{peak}}$ and the recombination times of both short and long trajectories on the same graph, illustrating the resemblance and relation between the different curves.

(a) *Squeezing dependent attosecond pulses (TDSE)*  (b) *Shift of peak moment of the attosecond pulse as a function of driver squeezing*

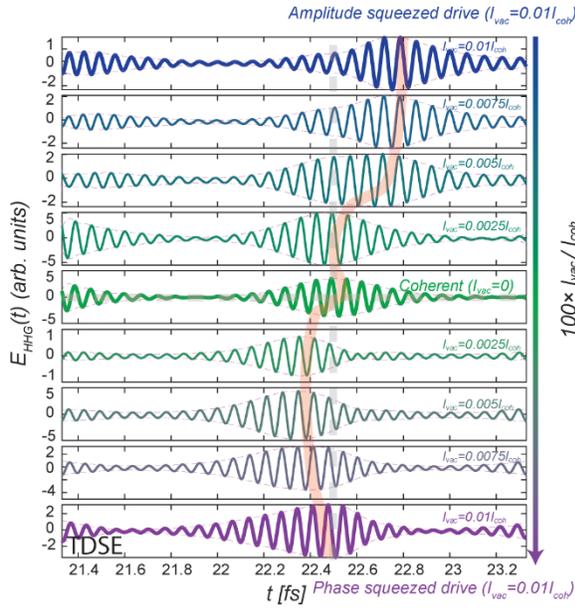
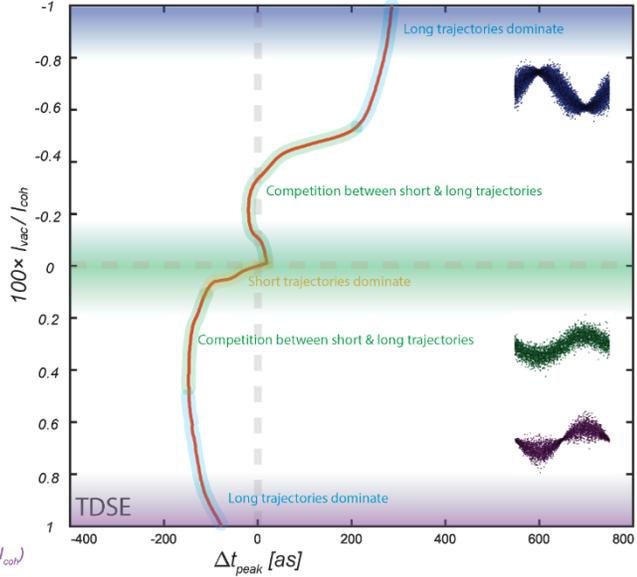

Competition between short and long trajectories as the origin of the temporal delay of the attosecond pulses (qSFA):

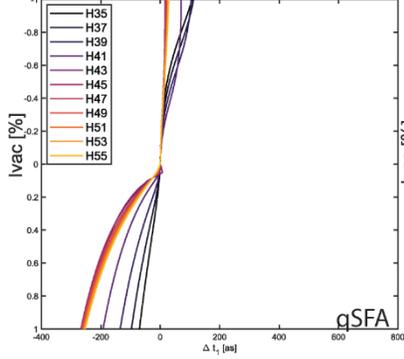
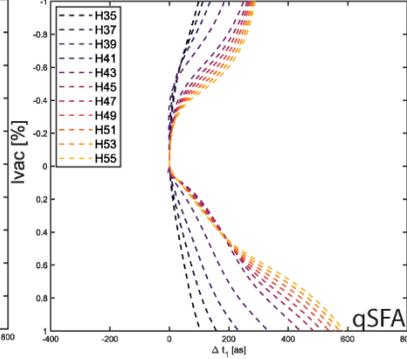
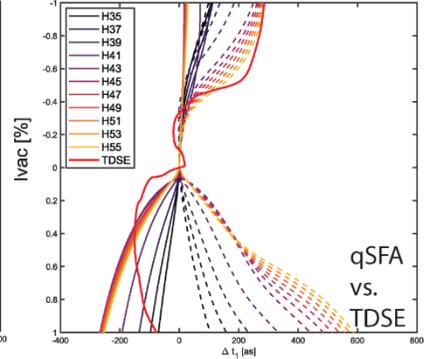

**Figure 3: The dependence of attosecond pulses on the squeezing of the driving field. (a)** Time-dependent Schrodinger equation (TDSE) simulations based on Eq.(7.a) provide the attosecond pulse shape for different levels of squeezing. **(b)** The peak time of the attosecond pulse $t_{peak}$ as a function of the squeezing strength $I_{vac}$ (negative $I_{vac}$ represents amplitude squeezing and positive represents phase squeezing). **(c)** Short trajectory recombination times as a function of squeezing for harmonics 35-55. **(d)** Long trajectory recombination times for harmonics 35-55. **(e)** Both $t_{peak}$ and recombination times displayed on the same graph, showing the competition between short and long trajectories as the origin of the complicated squeezing-dependence of $t_{peak} = t_{peak}(I_{vac})$.

## DISCUSSION

**Impact of vacuum fluctuations on high harmonic generation**

Even seemingly classical driving fields given by an electric field trace E(t) approximate Glauber coherent state[24], which is a quantum mechanical entity limited by the uncertainty principle and characterized by Poissonian photon statistics (Figure 4). Interestingly, deviations from classical results are always present because of vacuum fluctuations. Further investigation (SI section IV) shows that in a typical HHG experiment, the quantum optical correction to the recombination and ionization times is on the order of *yoctoseconds* ($10^{-24}$ sec), as shown in Figure 4(a-b).

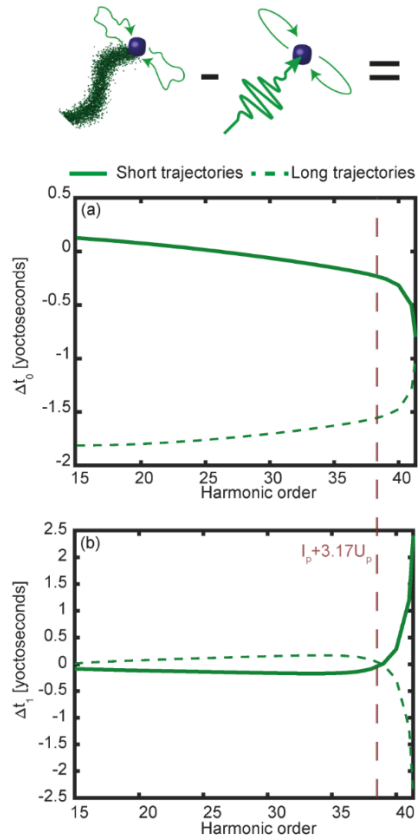

**Figure 4: Vacuum fluctuations induce yoctosecond time delays in electronic trajectories in high harmonic generation. (a,b)** Vacuum-fluctuations-induced corrections to ionization and recombination times, respectively. The calculation is described in detail in the SI, section IV.

**Photon-statistics force**

The previous sections established a relation between the photon statistics of the driving light and the dynamics of the driven system. This relation naturally hints at the existence of an effective force exerted on the electron by the photon statistics, and more specifically, by squeezing. Notably, the susceptibility of the dynamics to electric field fluctuations is due to the nonlinearity of the material system[47]. In this section, we show that indeed, for squeezed light, the trajectories obtained from the qSFA theory can be equivalently obtained from an effective semi-classical theory that includes an effective force added to the classical electric force of the driving pulse. We emphasize that while we derive an explicit formula only for squeezed light, an effective photon statistics force is a general phenomenon in strong-field physics.

To derive the photon statistics force for squeezed light, we begin with the quantum-optical action $S_q(\mathbf{p}, t', t, \alpha) = S_\alpha(\mathbf{p}, t', t) + i\log(P(\alpha))$ (Eq.(9)) with the appropriate $P(\alpha)$ distribution for squeezed coherent light. We then express the saddle point condition of qSFA: $\nabla_{\kappa_q}(S_q(\kappa_q) - \hbar\Omega t) = 0$ as 5 coupled algebraic equations in the parameters $\kappa_q = [p, t', t, \alpha_x, \alpha_y]$, where $\alpha_{x,y}$ are real-valued quadrature amplitudes. We proceed by solving $\partial_{\alpha_x} S_q = \partial_{\alpha_y} S_q = 0$ and are left with 3 equations $\partial_p S_q = \partial_t(S_q - \hbar\Omega t) = \partial_{t'} S_q = 0$. Finally, we integrate these equations to arrive at an effective semi-classical action $S_{\text{eff}}(\mathbf{p}, t', t)$, which exhibits the same saddle points as the quantum-optical action $S_q(\mathbf{p}, t', t, \alpha)$, and is given by

$$S_{\text{eff}}(p, t', t) = \int_{t'}^{t} dt'' \frac{\left[p - A_\gamma(t'') - \frac{i}{\omega^2} E_{\text{vac}}^2 \sin(\omega t'') \int_{t_0}^{t''} d\tau\, v(\tau) \sin(\omega \tau)\right]^2}{2}. \quad (010)$$

See sections IV and V in SI for detailed derivations. That is, the strong-field dynamics of the electron in the squeezed coherent field are formally equivalent to dynamics in an effective semi-classical theory, where the electron is subject to the classical vector potential $A_\gamma(t)$ and an effective photon statistics force which takes the form of a vector potential $A_{\text{sq}}(t)$:

$$A_{\text{sq}}(t'') = \frac{i}{\omega^2} E_{\text{vac}}^2 \sin(\omega t'') \int_{t_0}^{t''} d\tau\, v(\tau) \sin(\omega \tau). \quad (011)$$

Here, $E_{\text{vac}}$ is the amplitude of electric field fluctuations in the anti-squeezed quadrature, which in our case is the $\sin(\omega t)$ quadrature of the vector potential. The parameter $t_0$ is the initial moment of the dynamics and $v = v(\tau)$ is the velocity at each time. $A_{sq}(t'')$ depends on all times between $t_0$ and $t''$ (current time), that is, it expresses memory in the system.

With such a field in hand, we can derive Newtonian trajectories. Plugging $A_{sq}(t'')$ into Eq. (8.a) for the semi-classical dipole moment, we approximate the motion of the electron in the effective photon statistics force to be given by

$$z(t'') = e^{-\frac{E_{vac}^2}{\omega^2} v(t'') \sin(\omega t'') \int_{t_0}^{t''} d\tau v(\tau) \sin(\omega \tau)} \times z_{SC}(t). \tag{012}$$

$z_{SC}(t)$ is constructed out of the saddle point of the original (uncorrected) semi-classical SFA action (SC subscript for semi-classical). To obtain Newtonian trajectories, we replace $z_{SC}(t)$ with the classical trajectories $z_C(t)$ that solve Newton's 2$^{nd}$ law for an electron released at $t_0$ and driven by $A_\gamma(t)$. The resulting Newtonian trajectories of an electron in a coherent, phase squeezed, and amplitude squeezed states of light are displayed in Figure 5.(a-c). The dispersion of recombination energies for the different cases is plotted in Figure 5(d-f). As shown, the dispersion of recombination energies shows significant dependence on the photon statistics. Phase squeezed light results in a structurally similar dispersion curve to coherent state light, but with lower maximal recombination energy. Amplitude squeezed light results in maximal recombination energy that is similar to that of a coherent state, but with a significantly modified temporal structure.

**Conclusion**

We formulated a quantum optical theory of HHG, derived under the SFA. Our theory directly relates the driving fields' photon-statistics, the emitted radiation in the time domain, and the underlying electronic trajectories. We have shown that all these features of HHG are modified by the squeezing of the driver photonic state. From a fundamental perspective, our work reveals an effective force that holds the effect of any quantum photon statistics on the electron dynamics. The effect of squeezing on the electronic trajectories also modifies the time delays of the emitted attosecond pulses. This suggests that conventional techniques of attosecond science that resolve electronic trajectories could resolve the effect of the quantized fields. Looking forward, such techniques may hold the potential to unravel the role of light-matter entanglement in ultrafast material dynamics. Overall, we believe our work will play a central role in the emerging field of XUV quantum state engineering & tomography[34,48,49], potentially advancing quantum spectroscopy [48] to the strong-field regime.

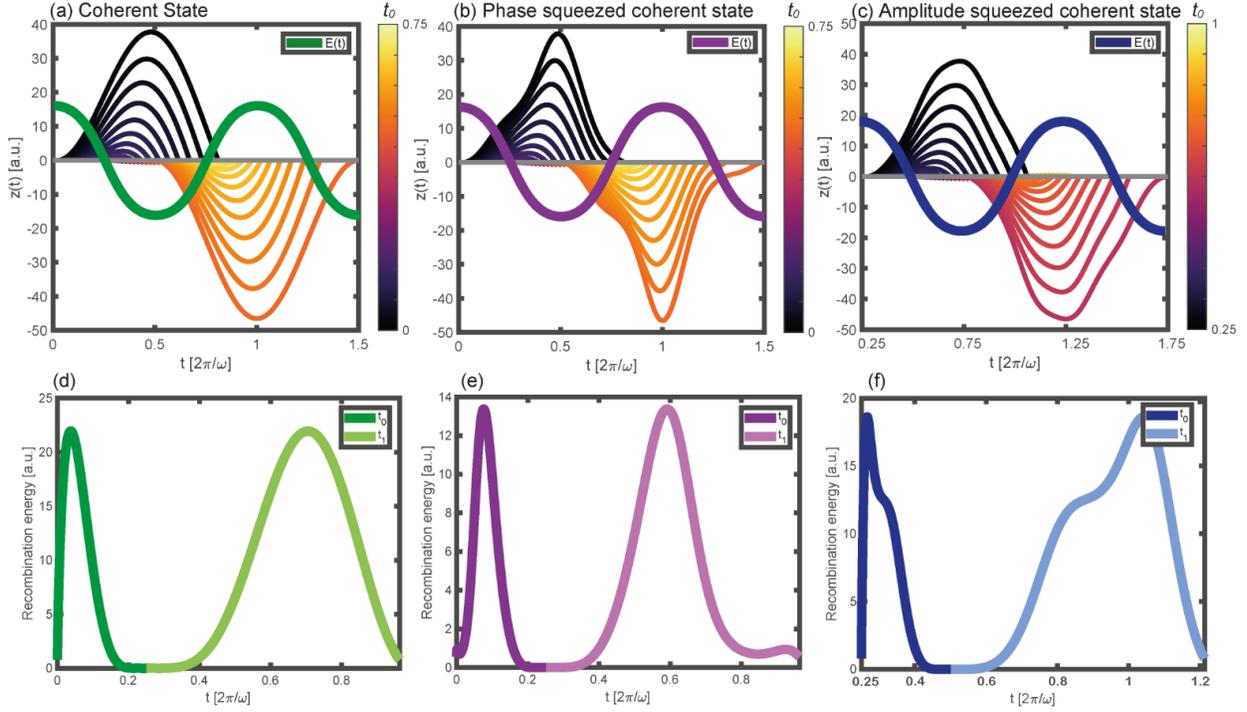

**Figure 5 The influence of the effective photon-statistics force: Newtonian trajectories. (a)** Classical Newtonian trajectories driven by a classical electromagnetic wave. **(b,c)** Newtonian trajectories driven by a phase-squeezed and an amplitude-squeezed coherent state of light, respectively, corresponding to the effective squeezing force. **(d,e,f)** Dispersion of recombination energies for an electron driven by **(d)** coherent, **(e)** phase-squeezed, and **(f)** amplitude-squeezed coherent states of light.

**Data Availability Statements**

The data supporting the findings of this study are available from the corresponding author upon reasonable request.

**Code Availability Statements**

The code supporting the findings of this study are available from the corresponding author upon reasonable request.

**Acknowledgments**

This work was supported by the Israel Science Foundation (Grant No. 1781/18). M.E.T. gratefully acknowledges the support of the Council for Higher Education scholarship for excellence in quantum science and technology & Helen Diller scholarship for excellence in quantum science & technology.

**Contributions**

**Competing interests**

The authors declare no competing interest.